\documentclass[twocolumn,aps]{revtex4}
 
\usepackage{amssymb} 
\usepackage{amsmath} 
\usepackage{amsbsy} 
\usepackage{bm} 
\usepackage{graphicx}

\begin{document}

\title{Modelling recorded crime: a full search for cointegrated models} 
\author{J.\ L.\ van Velsen} 
\affiliation{Dutch Ministry of Justice, Research and Documentation Centre (WODC), P.\ O.\ Box 20301, 2500 EH The Hague, The Netherlands}
\email{j.l.van.velsen@minjus.nl} 
\begin{abstract} A modelgenerator is developed that searches for cointegrated models among a potentially large group of candidate models. The generator employs the first step of the Engle-Granger procedure and orders cointegrated models according to the information criterions AIC and BIC. Assisted by the generator, a cointegrated relation is established between recorded violent crime in the Netherlands, the number of males aged 15-25 years (split into Western and non-Western background) and deflated consumption. In-sample forecasts reveal that the cointegrated model outperforms the best short-run models. 
\end{abstract} 

\maketitle 

\section{Introduction}
\label{introduction}

The statistical relation between recorded crime and economic and demographic variables, such as unemployment and the number of young males, is not only of fundamental importance \cite{hale91,britt94,greenberg01}, but becomes increasingly important in forecasting tools for policy and planning decisions \cite{harries03, deadman03}. As recorded crime and its predictor variables are generally integrated of order one, two kinds of models exist. There are short-run models specified in first differences and cointegrated models specified in levels. In a short-run model, crime has a strong stochastic dynamic of its own, eventually drifting away uncontrollably from the relation with its predictors. Conversely, in a cointegrated model, crime fluctuates with constant variability around the relation with its predictors, thereby preserving its importance in the long run. As a consequence, cointegrated models are to be favored over short-run models when making long term forecasts. Theoretically, given a set of predictor variables, one cannot argue conclusively whether a cointegrated model applies. Empirically, however, a time series analysis indicates whether a cointegrated model is supported by the data or not \cite{engle87,johansen88}. For example, no cointegration was found between crime and unemployment in England and Wales \cite{hale91}, while a cointegrated relation was established between robbery rate and divorce rate in the United States \cite{greenberg01}. 
      
The objective of this work is to go beyond a time series analysis for a fixed set of predictor variables. Rather, we consider all subsets of a set of potential predictor variables and search for cointegrated models. The kind of crime under investigation is violent crime in the Netherlands, composed of robbery, assault, homicide, rape and extortion. The set of potential predictor variables consists of unemployment, the number of males aged 15-25 years of Western background, the number of males aged 15-25 years of non-Western background, the number of divorces and deflated consumption. A manual inspection of all subsets is not feasible because the number of subsets, and thereby the number of candidate models, grows exponentially with the number of potential predictor variables. Even with five potential predictor variables and allowing for a potential error correction parameter, a constant or a constant and a trend, nearly two hundred candidate models exist. To investigate these models, we developed a modelgenerator that automatically builds and estimates them and subsequently checks for cointegration. The cointegration check consists of the first step of the Engle-Granger (EG)  procedure \cite{engle87}. The generator discards models that are not cointegrated and then estimates the cointegrated models using NLS regression. Finally, the generator orders the cointegrated models using the information criterions AIC \cite{akaike73} and BIC \cite{schwarz78}.

\section{Modelstructure and preliminary data analysis}
\label{modelstructure}

A necessary condition for the existence of a cointegrated relation is that all variables involved are integrated of the same non-vanishing order. ADF tests \cite{said84} indicate that recorded violent crime and its potential predictor variables do not reject one or more null hypotheses of a unit root up to the 0.10 level. These tests were performed on annual data ranging from 1978 to 2006 and the detailed results are listed in Table \ref{ADFKPSS}. We also performed KPSS tests \cite{kwiatkowski92} that reverse the null hypothesis (a unit root) and the alternative hypothesis (stationarity) of the ADF tests. The KPSS results are included in Table \ref{ADFKPSS}. The probabilities of $x_{2}$ having a unit root and of $x_{2}$ being stationary are all rather small (see Table \ref{ADFKPSS} for the symbolic notation of the variables). Further analysis reveals that $x_{2}$ is more likely to be integrated of order two than of order one. However, based on the joint picture emerging from Table \ref{ADFKPSS}, we conclude that all variables (including $x_{2}$) are integrated of order one. Having met the necessary condition, the following cointegrated model may exist
\begin{equation}
y_{t} = \sum_{i \in S} \beta_{i}x_{it}+\eta_{t}, \quad \mbox{where} \quad \eta_{t}=\frac{\epsilon_{t}}{1-\phi \mathcal{L}} \quad \mbox{and} \quad |\phi| < 1. 
\label{boxjenkins}
\end{equation}
Here, $t$ denotes discrete time measured in years, $S$ is a non-empty subset of the integer set $S_{0}=\{1,2,3,4,5\}$, $\beta_{i}$ is the coefficient connecting $x_{i}$ to $y$ and $\epsilon$ is Gaussian white noise. The operator $\mathcal{L}$ is the backshift operator retarding a variable with one year and the autoregressive noise $\eta$ is stationary and causal with correlation length $-1/\ln |\phi|$ measured in years. Model (\ref{boxjenkins}) may be augmented with a constant $c$ or a constant and a trend $c+\delta t$. 

\begingroup
\squeezetable
\begin{table}[h!]
\caption{Results of ADF tests and KPSS tests on recorded violent crime and its potential predictor variables. The lag length of the ADF tests is selected with BIC. The numbers in the ADF columns are the probabilities of the variable having a unit root. The subcolumns `0', `C' and `CT' correspond to a random walk around, respectively, zero, a constant and a constant and a trend. The probabilities are based on finite-sample distributions of the DF statistic \cite{dickey79} of Ref.\ \cite{mackinnon96}. The KPSS tests are performed with the Bartlett kernel and the data-based Newey-West bandwidth selection procedure \cite{newey94}. The numbers in the KPSS subcolumns `C' and `CT' are the probabilities of the variable being stationary around, respectively, a constant and a constant and a trend. These probabilities are based on the asymptotic test statistics of Ref.\ \cite{kwiatkowski92}.}
\begin{tabular}{|l|l||c|c|c|c|c|} \hline
\multicolumn{2}{|c||}{variable} & \multicolumn{3}{|c|}{ADF} & \multicolumn{2}{|c|}{KPSS} \\ \cline{1-7}
symbol & description & 0 & C & CT & C & CT \\ \hline
y    & recorded violent crime & 1.00 & 0.98 & 0.65 & $< 0.05$ & $< 0.05$ \\ \hline 
$x_{1}$ & unemployment (number  & 0.63 & 0.02 & 0.00 & $> 0.10 $ & $< 0.10$ \\ 
  & of persons aged 15-65 years &  &  &  &  &  \\ 
  & looking for a job) &  &  &  &  &  \\ \hline
$x_{2}$ & number of males  & 0.13 & 0.01 & 0.00 & $< 0.05$ & $\approx 0.10$  \\
  & aged 15-25 years &  &  &  &  &  \\ 
  & of Western background &  &  &  &  &  \\ \hline
$x_{3}$ & number of males  & 0.99 & 1.00 & 0.01 & $< 0.05$ & $< 0.10$  \\
  & aged 15-25 years &  &  &  &  &  \\ 
  & of non-Western background &  &  &  &  &  \\ \hline
$x_{4}$ & number of divorces & 0.83 & 0.07 & 0.39 & $< 0.10 $ & $> 0.10$ \\ \hline  
$x_{5}$  & deflated consumption  & 1.00 & 1.00 & 0.06 & $< 0.05 $ & $< 0.05$ \\ \hline  
\end{tabular}
\label{ADFKPSS}
\end{table}
\endgroup

The relation between the $x_{i}$ and $y$ in model (\ref{boxjenkins}) is instantaneous. This seems too restrictive as more complicated temporaneous relations are plausible. For example, in modelling the unemployment-crime relation, lagged unemployment has been argued to influence current crime numbers from a motivational perspective \cite{cantor85}. Indeed, such an effect has been found in a short-run model of violent crime (robbery excluded) \cite{hale91}. Mathematically, this corresponds to $\beta_{1}$ being a transfer function rather than a coefficient. The transfer function would be a polynomial in $\mathcal{L}$ corresponding to the substitution $\beta_{1} \rightarrow \beta_{10}+\beta_{11}\mathcal{L}$. Here, the coefficient $\beta_{10} < 0$ corresponds to the instantaneous effect of $x_{1}$ on $y$ argued from the opportunity perspective. The coefficient $\beta_{11} > 0$ corresponds to the lagged influence of $x_{1}$ on $y$ argued from the motivational perspective \footnote{The signs of $\beta_{10}$ and $\beta_{11}$ do not always agree with empirical findings. For example, in Ref. \cite{greenberg01}, several cointegrated models were tried, all indicating $\beta_{11} < 0$. In addition, the short-run model of robbery of Ref.\ \cite{hale91} has $\beta_{10} > 0$ and $\beta_{11} < 0$.}. To find out whether our historical data contain evidence of such a temporaneous relation, we employ the Box-Jenkins methodology \cite{box76}. This consists of stationarizing $x_{1}$ and $y$ by differencing, prewhitening $x_{1}$ with an ARMA filter, applying the filter to the stationarized $y$ and inspecting the correlogram of the prewhitened $x_{1}$ and the filtered $y$. The correlogram did not have any significant lag structure, indicating that $\beta_{1}$ is a scalar. Now we turn to the temporaneous relations between $y$ and the remaining predictor variables. The variables $x_{2}$ and $x_{3}$ are based on the age distribution of crime \cite{hirschi83, steffensmeier89}. Their impact on $y$ is instantaneous, but may also have a lagged component based on the following argument. A relatively violent (non-violent) group of individuals that is within the age cohort at time $t$, but outside the cohort at time $t' > t$, may influence $y_{t'}$ such that it is larger (smaller) than expected based on $x_{2t'}$ and $x_{3t'}$. That is, $\beta_{2}$ and $\beta_{3}$ may be transfer functions, and when they are, probably with a non-vanishing number of roots in their denominators to capture the long term correlations. We again employ the Box-Jenkins methodology and find that $\beta_{2}$ and $\beta_{3}$ are scalars. Finally, $x_{4}$ is indicative of strain in society \cite{greenberg01} and $x_{5}$ is important from both the motivational ($\beta_{5} < 0$) and the opportunity perspective ($\beta_{5} > 0$). Both variables are assumed to impact $y$ instantaneously, an assumption confirmed by a Box-Jenkins analysis.   

All variables are left untransformed. Taking the logaritm of $y$ and the $x_{i}$ carries the additive model (\ref{boxjenkins}) over into a multiplicative one. The transformed model with $S=\{j\}$ takes the form $y_{t}=x_{jt}^{\beta_{j}}e^{\eta_{t}}$, such that $y$ depends non-linearly on $x_{j}$ (if $\beta_{j} \neq 1$) and the variability of $y$ depends on $x_{j}$. Traces of these signatures were not found in scatter plots of $y$ and the $x_{i}$. Also, in the Box-Jenkins analyses, all variables became stationary (up to the 0.10 level of the KPSS test with a constant) after differencing without first taking the logaritm. This also holds for deflated consumption, which, in contrast to consumption, does not need to be transformed prior to differencing.

\section{The modelgenerator}
\label{modelgenerator}

The modelgenerator builds all models based on the $(2^{5}-1)$ non-empty subsets of $S_{0}$. Possibly augmented with a constant or a constant and a trend, there are $3(2^{5}-1)$ candidate models. In addition, the generator considers models with restriction $\eta=\epsilon$ (this corresponds to $\phi=0$) and models without this restriction. This brings the total number of candidate models to $6(2^{5}-1)=186$. The generator checks each model for cointegration and then orders the cointegrated models according to information criterions. These two stages of operation are described below. 

\subsection{Checking models}

On multiplying both sides of Eq.\ (\ref{boxjenkins}) by $(1-\phi \mathcal{L})$ and subtracting $(1-\phi)y_{t-1}$, the model takes error correction (EC) form
\begin{equation}
\Delta y_{t}= \sum_{i \in S} \beta_{i} \Delta x_{t} + (\phi-1) \left[ y_{t-1}-\sum_{i \in S} \beta_{i} x_{i(t-1)} \right] + \epsilon_{t}.
\label{errcor}
\end{equation}
Here, $\Delta=1-\mathcal{L}$ is the difference operator and the term with square brackets is the EC term. The regression equation (\ref{errcor}) is non-linear in its parameters $\beta_{i}$ (and possibly a constant or a constant and a trend) and error correction parameter $\phi$ \footnote{The constant $c$ and trend $\delta t$ that can be augmented to Eq.\ (\ref{boxjenkins}), carry over to the EC term. In addition, $\delta t$ appears as $\delta \phi$ outside the EC term.}. The first step of the EG procedure consists of estimating the coefficients $\beta_{i}$ (and possibly a constant or a constant and a trend) with the restriction $\phi=0$. Due to the restriction, the regression is linear and OLS is used. If a DF test on the residuals fails to reject the null hypothesis of a unit root up to the 0.05 level according to the asymptotic test statistics of Ref.\ \cite{mackinnon96}, the generator considers the two corresponding candidate models (the one with $\phi=0$ in the first place and the one without this restriction) not cointegrated and discards them. 

In the second step of the EG procedure, the term in square brackets in Eq.\ (\ref{errcor}) is replaced by the lagged residuals of the first step and the resulting equation is estimated by OLS. As computation time is not really an issue with a moderate number of candidate models, the generator does not employ the second EG step and rather solves Eq.\ (\ref{errcor}) directly using a Marquardt NLS algorithm. The algorithm is assisted by analytic derivatives of the  squared residuals to $\phi$ and executed with a coefficient accuracy of $1 \cdot 10^{-4}$ and a maximum number of $5 \cdot 10^{2}$ iterations. After estimating the parameters, the generator inspects the residuals and discards all models that reject the null hypothesis of no serial correlation up to the 0.20 level of the Breusch-Godfrey (BG) LM test \cite{breusch78, godfrey78}. The BG LM test is performed with two lags and pre-sample residuals are set to zero.       

\subsection{Ordering models}

The information criterions AIC and BIC take the form
\begin{equation}
{\rm AIC}=\ln {\rm SSR} + \frac{2N}{n}, \quad {\rm BIC}=\ln {\rm SSR} + \frac{N\ln n}{n}.
\end{equation}
Here, SSR is the sum of squared residuals of the NLS estimation, $N$ is the total number of estimated parameters and $n$ is the sample size. Models with smaller AIC -- an estimator of the cross-entropy between the unknown operating model and the candidate model, averaged over the operating model -- are favored over models with larger AIC. The same holds for BIC which is related to the posterior probability of a model as $\sim {\rm exp}(-{\rm BIC}/2)$. In finite samples, AIC underestimates the averaged cross-entropy and favors models that are too complex \cite{hurvich89}. If $n \ge 8$ such that $\ln n > 2$, BIC has a larger penalty term than AIC and favors simpler models. Which of the two criterions is more appropriate for finite $n \ge 8$ cannot be said beforehand and the generator employs both.   

\section{Results}
\label{results}

Out of the 186 candidate models, the generator produces 15 cointegrated models. The top-6 of best models according to AIC happens to coincide with that according to BIC in the sense that they contain the same models, but with a different ordering. These 6 models, as well as the corresponding evidence ratio's (ER's), are listed in Table \ref{cointmodels}. ER is related to AIC and BIC as, respectively, ${\rm exp}(-({\rm AIC}-{\rm AIC}_{\rm min})/2)$ and ${\rm exp}(-({\rm BIC}-{\rm BIC}_{\rm min})/2)$. Here, ${\rm AIC}_{\rm min}$ (${\rm BIC}_{\rm min}$) denotes the smallest AIC (BIC) out of the cointegrated models. For BIC, ER measures the ratio of a posterior probability and the highest posterior probability. The results of Table \ref{cointmodels} indicate that all 6 models contain the number of non-Western males $x_{3}$ and deflated consumption $x_{5}$. In contrast, the number of divorces $x_{4}$ does not appear in any of the cointegrated models. 

In an additional run of the generator, we used $(x_{2}+x_{3})$ rather than $x_{2}$ and $x_{3}$. The reason for this is the following. Suppose the unknown operating model contains the number of males aged 15-25 years without distinguishing between Western and non-Western background. Then, models containing $x_{2}$ and $x_{3}$ unnecessarily require estimation of both $\beta_{2}$ and $\beta_{3}$, while in fact $\beta_{2}=\beta_{3}$. Out of the $6\cdot 2^{3}=48$ candidate models containing $(x_{2}+x_{3})$, two are found cointegrated. Both models have a larger AIC and BIC than models I-VI and leave the results of Table \ref{cointmodels} unaffected.

\begingroup
\squeezetable
\begin{table}[h!]
\caption{Schematic listing of cointegrated models. The variables $x_{i}$ represent the potential predictor variables as listed in Table \ref{ADFKPSS}. The symbol `X' indicates inclusion of the corresponding predictor variable (or constant $c$, trend $t$ or error correction parameter $\phi$). The columns `ER AIC' and `ER BIC' hold, respectively, the ER's according to AIC and BIC. While the models are ordered according to BIC, the numbers in brackets in the column `ER AIC' denote their order according to AIC. The numbers in the column `BG LM' are the probabilities of the residuals having no serial correlation according to the BG LM test. The column `EG DF' holds the DF statistics of the OLS residuals of the first step of the EG procedure and the numbers in brackets are the corresponding asymptotic test statistics at the 0.05 level of Ref.\ \cite{mackinnon96}.}
\begin{tabular}{|l||c|c|c|c|c|c|c|c||c|c|c|c|} \hline
\multicolumn{9}{|c|}{model} & ER  & ER & BG  & EG DF \\ \cline{1-9}
 number &   $c$ & $t$ & $x_{1}$ & $x_{2}$ & $x_{3}$ & $x_{4}$ & $x_{5}$ & $\phi$ &  AIC & BIC & LM &  \\ \hline
 I &    X &   &   &   & X &   & X &  & 0.99 (3)  & 1.00   & 0.38 & -3.92 (-3.74) \\ \hline
 II      &  X &   &   & X  & X &   & X &   & 1.00 (2) & 0.99   & 0.37 & -4.41 (-4.10) \\ \hline
 III      &  X    & X   & X       &       & X       &    & X &   & 1.00 (1) & 0.97   & 0.35 & -4.67 (-4.43) \\ \hline
 IV &    X &   &  &  & X  &   & X & X & 0.96 (6) & 0.95   & 0.84 & -3.92 (-3.74) \\ \hline
 V &    X & X  &   & X & X &   & X &  & 0.98 (4) & 0.94   & 0.35 & -4.48 (-4.43) \\ \hline
 VI &    X &  & X & X  & X  &   & X &   & 0.97 (5) & 0.94   & 0.37 & -4.49 (-4.41) \\ \hline
\end{tabular}
\label{cointmodels}
\end{table}
\endgroup

To investigate the cointegration structure of models I-VI, we consider the VEC (vector error correction) models 
\begin{equation}
\Delta z_{t} = \left\{ \begin{array}{ll} \alpha\left[\gamma + \beta^{\rm T} z_{t-1}\right] + \mu_{t} & \quad \mbox{(a)} \\
\alpha \left[ \gamma + \delta t+\beta^{\rm T}z_{t-1} \right] + \alpha_{\perp} \delta' + \mu_{t}. & \quad \mbox{(b)} \end{array} \right.         
\label{vec}
\end{equation}
Here, $z$ is the 5-dimensional columnvector $(y,x_{1},x_{2},x_{3},x_{5})^{\rm T}$ and `${\rm T}$' denotes the transpose of a matrix or vector. The matrices $\alpha$ and $\beta$ are $5 \times r$ matrices, where $r \le 4$ is the cointegration rank. The 5-dimensional columnvector $\mu$ is Gaussian white noise with $5 \times 5$ covariance matrix $\Omega$. The $r$-dimensional columnvector $\gamma$ holds constants within the EC terms $\alpha\left[\gamma + \beta^{\rm T} z_{t-1}\right]$ and $\alpha \left[ \gamma + \delta t+\beta^{\rm T}z_{t-1} \right]$. Similarly, the $r$-dimensional columnvector $\delta$ holds prefactors of the trend $t$ inside the EC term $\alpha \left[ \gamma + \delta t+\beta^{\rm T}z_{t-1} \right]$. The $(5-r)$-dimensional columnvector $\delta'$ corresponds to trends in the columnspace of the $5 \times (5-r)$ matrix $\alpha_{\perp}$ satisfying $\alpha^{\rm T}\alpha_{\perp}=0$. The difference between the VEC model (\ref{vec}) and the EC model (\ref{errcor}) is that the latter describes the dynamics of $y$ only, while the former describes the joint dynamics of $y$ and $(x_{1},x_{2}, x_{3},x_{5})$. The correspondence between the VEC model and the EC model is, that, conditioned on $(x_{1},x_{2}, x_{3},x_{5})$, the stochastic process of $y$ implied by the VEC model takes the form of the EC model \footnote{Rephrased in terms of the noise process $\eta$ and transfer functions $\beta_{i}(\mathcal{L})$, the conditional dynamics of $y$ implied by (\ref{vec}) is more general than (\ref{errcor}) as it allows for a rational $\beta_{i}(\mathcal{L})$ with numerator $(\beta_{i0}+\beta_{i1}\mathcal{L})$ and denominator $(1-\phi\mathcal{L})$. In the EC model, this corresponds to different coefficients inside and outside the EC term.}.  

We first consider models I, II, IV, VI and perform a Johansen cointegration test \cite{johansen88} to find $r$, $\alpha$, $\beta$ and $\gamma$ of VEC model (a) without trends. Up to the 0.05 level of Ref.\ \cite{mackinnon99}, the trace test indicates $r=3$ (LR = 0.065) and the maximum eigenvalue test indicates $r=4$ (LR = 0.022). We accept $r=4$. This means that there is only one common non-stationary signal and the 4 linear combinations corresponding to the elements of $\beta^{\rm T}z$ are such that this signal is cancelled. To see if this agrees with an EC model, we try to find coefficients $\xi_{i}$ such that $\sum_{i=1}^{4}\xi_{i}^{\vphantom{\rm T}}(\pmb{\beta}_{i}^{\rm T},\gamma_{i}^{\vphantom{\rm T}})(z^{\rm T},c)^{\rm T}$ equals the EC term of the EC model. Here, $\pmb{\beta}_{i}$ denotes the $i$-th column of $\beta$. Indeed, for each EC model, a set of coefficients $\{\xi_{i}\}$ exists within error bounds of a standard deviation above and below the estimated parameters. Now we turn to models III, V and find $r$, $\alpha$, $\beta$, $\gamma$ and $\delta$ of VEC model (b) with trends. Up to the 0.05 level, both the trace test and the maximum eigenvalue test indicate $r=3$, with, respectively, LR = 0.079 and LR = 0.063. In this case, we search for coefficients $\xi_{i}$ such that $\sum_{i=1}^{3} \xi_{i}^{\vphantom{\rm T}}(\pmb{\beta}_{i}^{\rm T},\gamma_{i}^{\vphantom{\rm T}},\delta_{i}^{\vphantom{\rm T}})(z^{\rm T},c,t)^{\rm T}$ equals the EC term of an EC model. It turns out that, for both models, sets of coefficients $\{\xi_{i}\}$ do not exist within error bounds. In other words, models III, V are not consistent with the VEC model at the 0.05 level and we discard them.

The estimated parameters of models I, II, IV, VI are listed in Table \ref{cointest}. The estimated parameter of $x_{1}$ in model VI is insignificant. The same holds for the EC parameter $\phi$ in model IV. As it does not hold $x_{2}$, model I is less plausible than model II. We conclude that model II is the best cointegrated model.

\begin{table}[h!]
\caption{Estimated parameters and t-ratio's (in brackets) of models I, II, IV, VI of Table \ref{cointmodels}.}
\begin{tabular}{|c||c|c|c|c|c|c|} \hline 
\multicolumn{7}{|c|}{model} \\ \cline{1-7}
number &   $c \cdot 10^{3}$ & $x_{1} \cdot 10^{-3}$ & $x_{2}\cdot 10^{-2}$ & $x_{3}\cdot 10^{-2}$ & $x_{5}$ \footnote{expressed in millions of euros deflated with respect to the year 2000} &  $\phi$   \\ \hline
I &  $-59$             &      &                        & $31.1$               & $0.37$    &              \\ 
  &  $(-10.2)$         &      &                        & $(5.3)$              & $(8.0)$   &                 \\ \hline
II & $-89$         &    & $1.5$                         & $28.3$               & $0.44$    &              \\ 
   &  $(-4.6)$    &      & $(1.6)$                      & $(4.7) $             & $(7.0)$   &              \\ \hline
IV &  $-57$             &       &                       & $33.2$               & $0.35$    & $0.26$       \\ 
    &  $(-7.2)$          &      &                        & $(3.9) $             & $(5.4)$   & $(1.2)$         \\ \hline
VI &  $-95$          &  3.2 & $1.5$                        & $23.7$               & $0.47$    &        \\ 
  &  $(-4.0)$     &  (0.5)   & $(1.6)$                     & $(2.1)$              & $(4.8)$   &          \\ \hline
\end{tabular}
\label{cointest}
\end{table}

\section{A comparison with short-run models}
\label{shortrunmodels}

In Sec.\ \ref{results}, only 15 cointegrated models were found out of 186 candidate models specified in levels. This means that for many subsets $S$, the proper model specification is in first differences rather than in levels. We therefore consider the short-run model 
\begin{equation}
\Delta y_{t}=\sum_{i \in S} \beta_{i} \Delta x_{it} + \epsilon_{t}.
\label{shortrun}
\end{equation}
Model (\ref{shortrun}) may be augmented with a constant and when it is, $S$ may the empty set. The generator is used to find the best out of the $2^{6}-1=63$ short-run models. (The cointegration check is switched off and the models are estimated with OLS.) The generator finds 51 models with a BG LM probability of no serial correlation exceeding 0.20. The top-5 models according to BIC are listed in Table \ref{shortrunest}. Models I and IV are equal up to an insignificant constant in model IV. Model II holds only $\Delta x_{3}$ and is considered too simple. The coefficient of $\Delta x_{1}$ in model V is insignificant and we conclude that models I and III are the best short-run models. 

In an additional run of the generator, we used $\Delta(x_{2} + x_{3})$ rather than $\Delta x_{2}$ and $\Delta x_{3}$. Out of the 16 candidate models containing $\Delta(x_{2} + x_{3})$, the generator selects 14 models. All these models have a larger AIC and BIC than models I-V and leave the results of Table \ref{shortrunest} unaffected.   

\begingroup
\squeezetable
\begin{table}[h!]
\caption{Estimated parameters and t-ratio's (in brackets) of the top-5 short-run models according to BIC. The columns `ER AIC' and `ER BIC' hold, respectively, the ER's according to AIC and BIC. The numbers in brackets in the column `ER AIC' denote the ranking according to AIC. The column `BG LM' holds the probabilities of the residuals having no serial correlation according to the BG LM test.}
\begin{tabular}{|c||c|c|c|c|c|c||c|c|c|} \hline 
\multicolumn{7}{|c||}{model} & ER & ER & BG  \\ \cline{1-7}
number & $c$       & $\Delta x_{1} \!\! \cdot \!\! 10^{-3}$ & $\Delta x_{2}$ & $\Delta x_{3}$ & $\Delta x_{4}$ & $\Delta x_{5}$ & AIC & BIC & LM  \\ \hline
I &          &                &                & 0.36           & 0.46           & 0.29   & 1.00 \!\! (1) & 1.00 & 0.39           \\
  &          &                &                & (2.9)          & (2.0)          & (2.7)  &  &  &     \\ \hline
II &       &         &           & 0.62 &  & & 0.91 \!\!\! (18) & 0.95 & 1.00 \\ 
   &       &         &          & (6.2) &  & &  &  & \\ \hline  
III  &       &  -17  &   & 0.65 & 0.48 & & 0.95 \!\! (5) & 0.95 & 1.00 \\     
   &       & (-2.0)  &    & (6.7) & (2.0) &  &  &  & \\ \hline   
IV & -464     &                &                & 0.41           & 0.48           & 0.32  & 0.97 \!\! (2) & 0.95 &    0.25        \\ 
  & (-0.4)   &                &                & (2.3)          & (2.0)          & (2.3)  &  &  &            \\ \hline
V &          & $-3.2$ &            & 0.39           & 0.47           & 0.26  & 0.97 \!\! (3) &  0.94 &  0.39         \\ 
   &        & (-0.3)         &                & (2.2)          & (2.0)          & (1.7)  &  &  &          \\ \hline
\end{tabular}
\label{shortrunest}
\end{table}
\endgroup
 
\begin{figure}[h!]
\includegraphics[width=9cm]{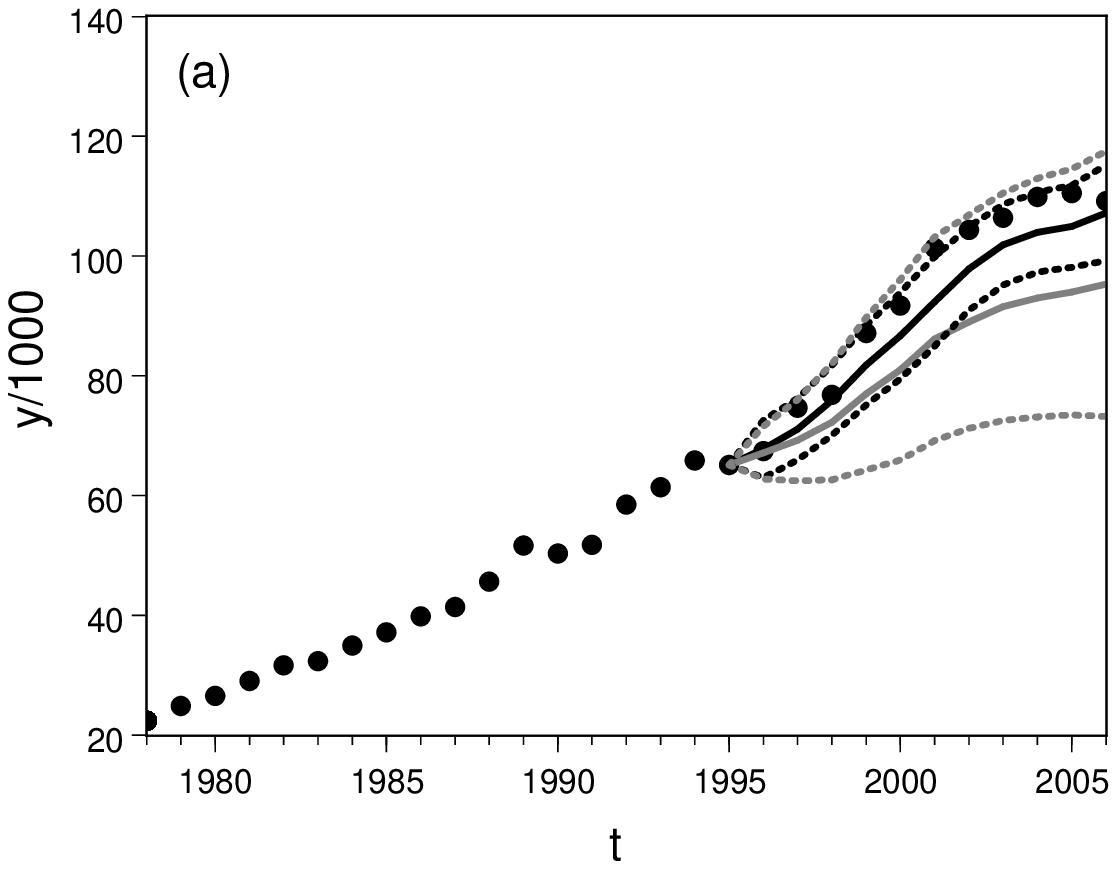}
\includegraphics[width=9cm]{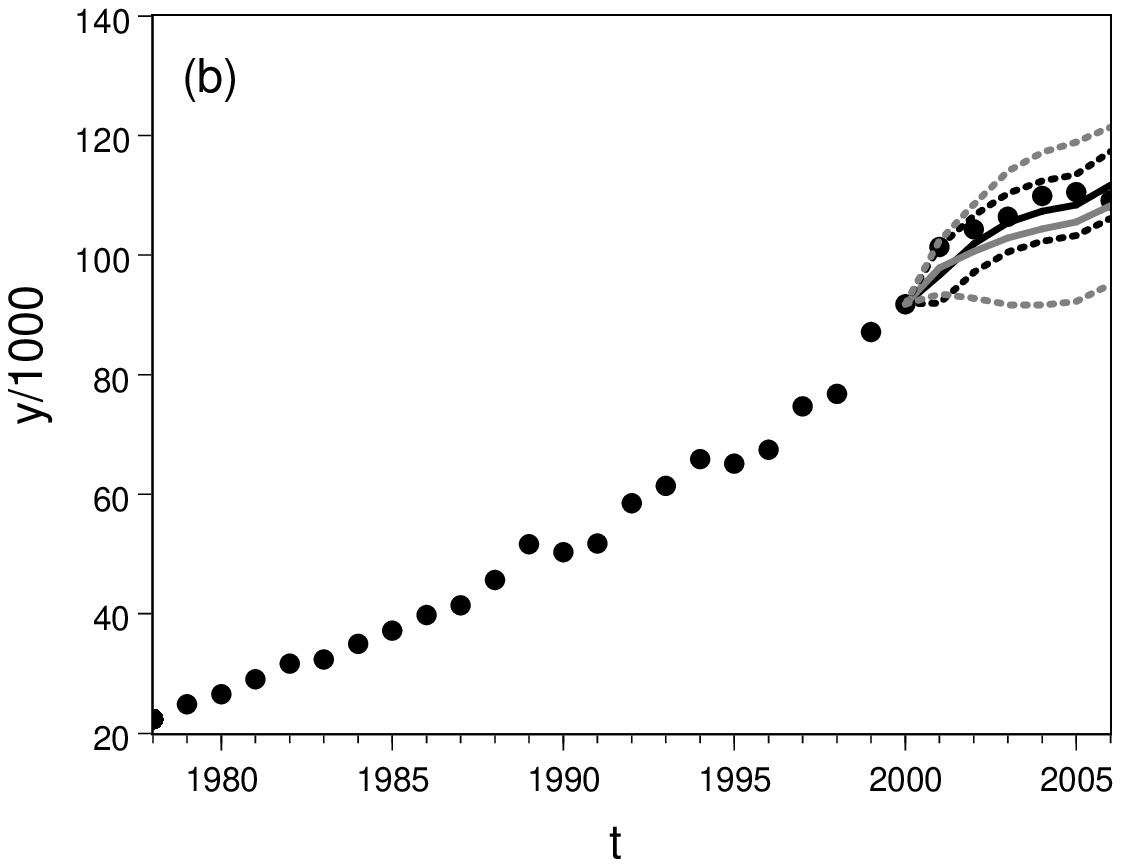}
\caption{Realizations (black dots) and mean forecasts of cointegrated model II of Sec.\ \ref{results} (solid black lines) and short-run model I of Sec.\ \ref{shortrunmodels} (solid gray lines). The dashed black (gray) lines constitute error bounds of two standard deviations above and below the mean forecasts of the cointegrated (short-run) model. In panel (a), the models are estimated based on realizations from 1978 up to 1995 and  forecasted from 1996 up to 2006. In panel (b), the models are estimated based on realizations from 1978 up to 2000 and forecasted from 2001 up to 2006.} 
\label{forecasts}
\end{figure}

The predictive power of short-run models I and III is compared to that of cointegrated model II by in-sample forecasts. The outcome of the comparison is not known beforehand: $x_{4}$ may be a good predictor of $y$ and the short-run models can outperform the cointegrated model when making short term forecasts. We make a long (short) term forecast, estimating the models using data from 1978 up to 1995 (2000) and forecasting them from 1996 (2001) to 2006. The forecasts are based on a Monte Carlo simulation of $10^{4}$ repetitions (including coefficient uncertainty) and consist of mean forecasts and error bounds. The results for short-run model I and the cointegrated model are indicated in Fig.\ \ref{forecasts}. Panel (a) holds the long term forecasts and indicates that the cointegrated model outperforms the short-run model in the sense that its mean predictions are closer to the realizations and that its error bounds are smaller. In panel (b), holding the short term forecasts, the mean forecasts of the models are very similar, but the error bounds of the short-run model are larger. The forecasts of short-run model III are similar to that of short-run model I. In panel (a), short-run model III would produce mean forecasts a bit worse than that of short-run model I. In panel (b), it would produce mean forecasts a bit closer to the realizations than that of short-run model I and the cointegrated model. The error bounds of short-run model III are comparable to that of short-run model I.

\section{Conclusions}
\label{conclusions}

We have developed a modelgenerator that searches for cointegrated models among candidate models specified in levels.  The generator employs the first step of the EG procedure to decide whether a model is cointegrated or not. Subsequently, it estimates the cointegrated models with NLS regression and checks the residuals with the BG LM test. Finally, it orders models according to both AIC and BIC. The generator is applied to recorded violent crime and five potential predictor variables: unemployment, the number of males aged 15-25 years of Western background, the number of males aged 15-25 years of non-Western background, the number of divorces and deflated consumption. The pattern emerging from the ordered list of cointegrated models is that the number of males aged 15-25 years of non-Western background and deflated consumption are the key predictors of recorded violent crime. Based on an analysis of VEC models, a plausibility argument and the statistical significance of estimated parameters, we select a cointegrated model out of the ordered list. The selected model holds, in addition to the key predictors, the number of males aged 15-25 years of Western background.
  
With the cointegration check switched off, the generator is used to find short-run models specified in first differences. The key predictors of the short-run models are the number of males aged 15-25 years of non-Western background and the number of divorces. Based on a plausibility argument and the statistical significance of estimated parameters, two short-run models are selected from the ordered list. In addition to the key predictors, one of the short-run models holds deflated consumption and the other unemployment. In-sample forecasts reveal that the cointegrated model outperforms the two short-run models in both mean predictions (long term) and error bounds (long term and short term). 
     
\end{document}